# Effect of surface treatment on vibration energy transfer of ultrasonic sonotrode

Zhang Xiaoyu[a,b], Zhang Lihua[b,c], Dong Fang[b,c], Jiang Ripeng[b,c], Zhang Yun[a,b]

a College of Mechanical and Electrical Engineering, Central South University, Changsha 410083, China

b State Key Laboratory of High Performance Complex Manufacturing, Changsha 410083, China

c Light Alloy Research Institute, Central South University, Changsha 410083, China

Corresponding author:

*Lihua Zhang;E-mail address:zhanglihua@csu.edu.cn

## Abstract

In this paper, two kinds of ultrasonic radiation rod with surface treatment (ion nitriding and vacuum carburizing) are selected to carry out finite element analysis on ultrasonic vibration system and casting system, and explore the influence of surface treatment on vibration energy transmission of radiation rod. The cavitation field of radiation rod with different surface treatment in water was obtained through the cavitation erosion area of aluminum foil in water by using the aluminum foil cavitation experiment, so as to verify the simulation results of sound pressure field in aluminum melt. The results show that the surface treatment weakens the vibration response of the radiating rod, reduces the longitudinal amplitude of the radiating rod, and reduces the amplitude of sound pressure transmitted into the aluminum melt.

## 1.Introduction

Ultrasonic-assisted casting can effectively refine grain size, purify melt, weaken alloying element enrichment, improve solute element segregation, increase solute element solid solubility, change crystalline phase distribution and other effects [1-4], thus improving ingot quality. The direct contact between the radiation rod and the aluminum melt will transfer the ultrasonic power to the aluminum melt. Cavitation corrosion and chemical corrosion caused by high temperature lead to the corrosion damage of the ultrasonic radiation rod, which reduces the service life of the radiation rod and affects the transmission efficiency of sound power, thus leading to the difficult quality of solidified tissue to meet the production requirements [5-7].

The corrosion can be reduced by treating the surface of the ultrasonic radiation rod. Liu Zhe etal[8]. formed a hard layer on the surface of the radiation rod in two ways of vacuum carburizing and ion nitriding, and analyzed and tested the corrosion resistance of the coating. They found that the anti-corrosion performance of the ultrasonic radiation rod after surface treatment was significantly improved. C.H.Tan et al[9]. used laser cladding to reorganize the microstructure of the matrix and improve the cavitation corrosion resistance of the Mn-Ni-Al Bronze alloy matrix. Liao Huali et al[10]. studied the vibration energy transmission performance of ultrasonic radiation rod theoretically by applying the corresponding formula, but only explored it theoretically without further verification.

Surface treatment can effectively slow down the corrosion damage of radiation rod and prolong the service life of radiation rod. However, the physical parameters such as the structure, elastic modulus and hardness of the radiation rod changed after surface treatment, and the vibration conduction law also changed accordingly. At present, the influence mechanism of surface treatment on ultrasonic radiation rod is not clear, and the influence on solidification structure of ingot after ultrasonic casting needs to be further explored. In this paper, the modal analysis and harmonic response analysis of three kinds of ultrasonic radiation rods were carried out by using finite element analysis, and the change of physical parameters of ultrasonic radiation rods after surface treatment on ultrasonic vibration was analyzed.

## 2.Dynamic characteristics of ultrasonic vibration system and simulation of sound pressure field in aluminum melt.

### 2.1 Dynamic simulation of ultrasonic vibration system

The structure diagram of ultrasonic vibration system is shown in Figure 1. The system includes: sandwich type transducer, amplitude transformer and radiation rod. The sandwich type transducer is composed of front and rear end caps, piezoelectric ceramic stack and fastening bolts. The ultrasonic transducer converts the electrical signal into the vibration signal of the whole vibration system by using the inverse piezoelectric effect of the piezoelectric ceramic sheet, and amplifies the amplitude of the vibration signal through the amplitude transformer, and finally transmits the vibration signal to the metal melt through the radiation rod, and the vibration of the radiation rod forms the sound field in the metal melt. In this section, three kinds of ultrasonic vibration system and ultrasonic casting system are modeled respectively. Using the coupling of ANSYS piezoelectric structure, modal analysis and harmonious response analysis are carried out respectively to obtain the dynamic characteristics of ultrasonic radiation rod.

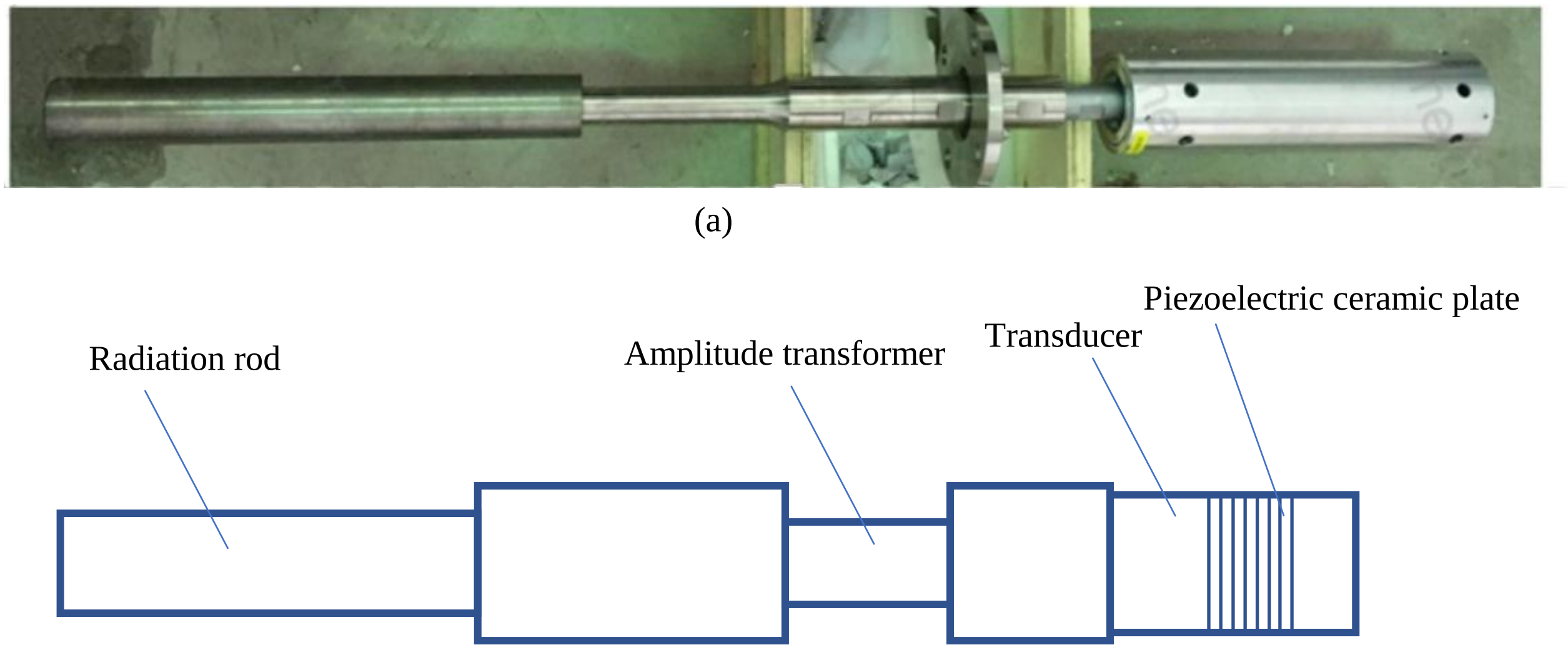

**Fig. 1.** Physical object and structure diagram of ultrasonic vibration system

(1) Nano indentation detection

Before the simulation, the structure parameters and physical parameters of the ultrasonic vibration system should be defined. Titanium alloy ultrasonic radiation rod ∅50×350 mm, select an ultrasonic radiation rod with a hard layer thickness of 55 um on the surface of the radiation rod after ion nitriding treatment, and also choose an ultrasonic radiation rod with a hard layer thickness of 53 um on the surface of the radiation rod after vacuum carburizing.

The physical parameters related to radiation rods can be detected by nano indentation. By obtaining the load and displacement data of the device's indentation head, a mechanical model is established to analyze the loading and unloading slope, and thus physical parameters such as elastic modulus, Poisson's ratio and hardness of three kinds of ultrasonic radiation rods are calculated [11-12]. The test results are shown in Table 1.

**Table 1**

Nano indentation experimental results of three radiation rods

| | Modulus of elasticity /GPa | Poisson's ratio | H ardness /GPa |
|---|---|---|---|
| Untreated | 115 | 0.34 | 3.3 |
| Ion nitriding | 144 | 0.33 | 6.5 |
| Vacuum carburizing | 130 | 0.22 | 4.1 |

According to the test results, the elastic modulus and hardness of the ultrasonic radiation rod increased after the surface treatment, and the elastic modulus and hardness of the ultrasonic radiation rod were the largest after the ion nitriding treatment. Compared with untreated radiation rod, the elastic modulus of treated radiation rod with ion nitriding and vacuum carburizing is increased by 25% and 13%, and the hardness is increased by 96% and 24%, respectively.

(2) Establishment of finite element model of ultrasonic vibration system

The main materials of the ultrasonic vibration system include aluminum alloy, 45 steel, PZT-8 type piezoelectric ceramics and titanium alloy. In addition to the existing data detected in the nano-indentation experiment, as shown in Table 1, the physical parameters of other materials are shown in Table 2. After the surface treatment of ultrasonic radiation rod, it is necessary to set the hard layer and titanium alloy substrate respectively, and set its connection mode as frictionless binding.

**Table 2**

Parametric characteristics of the three radiation rods

| Material | Modulus of elasticity /GPa | Density /kg.m$^{-3}$ | Poisson's ratio | Sound velocity /m.s$^{-1}$ |
|---|---|---|---|---|
| PZT8 | 9.1 | 7648 | — | 3465.3 |
| Titanium alloy | 110 | 4500 | 0.34 | 4921.1 |
| Structural steel | 21 | 7868 | 0.29 | 5188.4 |
| Aluminum melt | — | 2460 | — | 2282.0 |

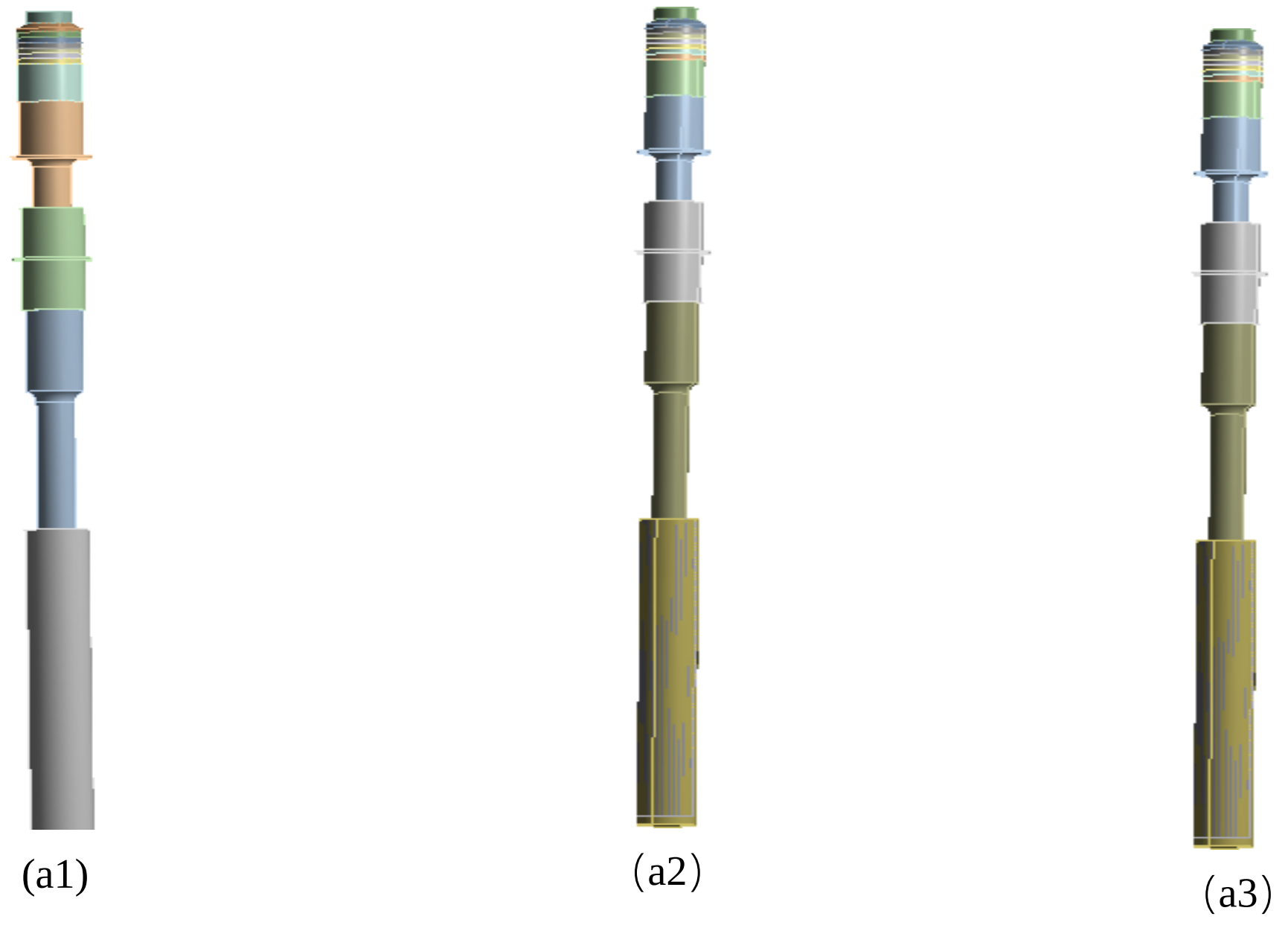

(a1)-(a3) Untreated, vacuum carburizing,ion nitriding

**Fig. 2.**Models of three radiation rods

The model diagram of the ultrasonic vibration system under the ultrasonic radiation rod with different surface treatments (untreated, vacuum carburizing and ion nitriding) is shown in Figure 2-(a1-a3). The finite element model can only be generated after meshing the model. The existence of hard layer on the surface of radiation rod after surface treatment requires meshing between the hard layer and the substrate. In this paper, ANSYS MESHING pretreatment software is used to divide the meshing of the ultrasonic vibration system. When dividing the meshing of the radiation rod without surface treatment, the mesh size is not set to 1mm. After calculation, it can be concluded that the number of mesh cells is 4101761 and the number of nodes is 11563583. Since the hard layer is only in micron magnitude, on the premise of not affecting the calculation accuracy, it is set as shell element for grid division. The contact mode between the hard layer and the matrix is set as binding contact, and the grid size is 0.1mm. After calculation, the number of grid elements is 5811865 and the number of nodes is 23463694.

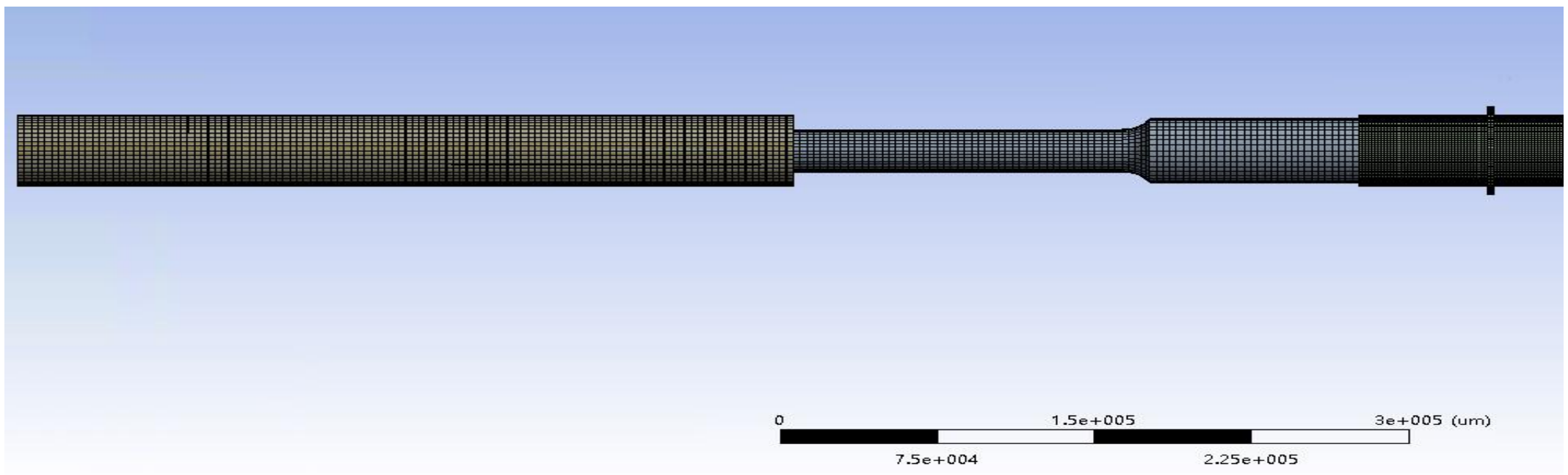

(a) Grid division diagram of untreated radiation rod

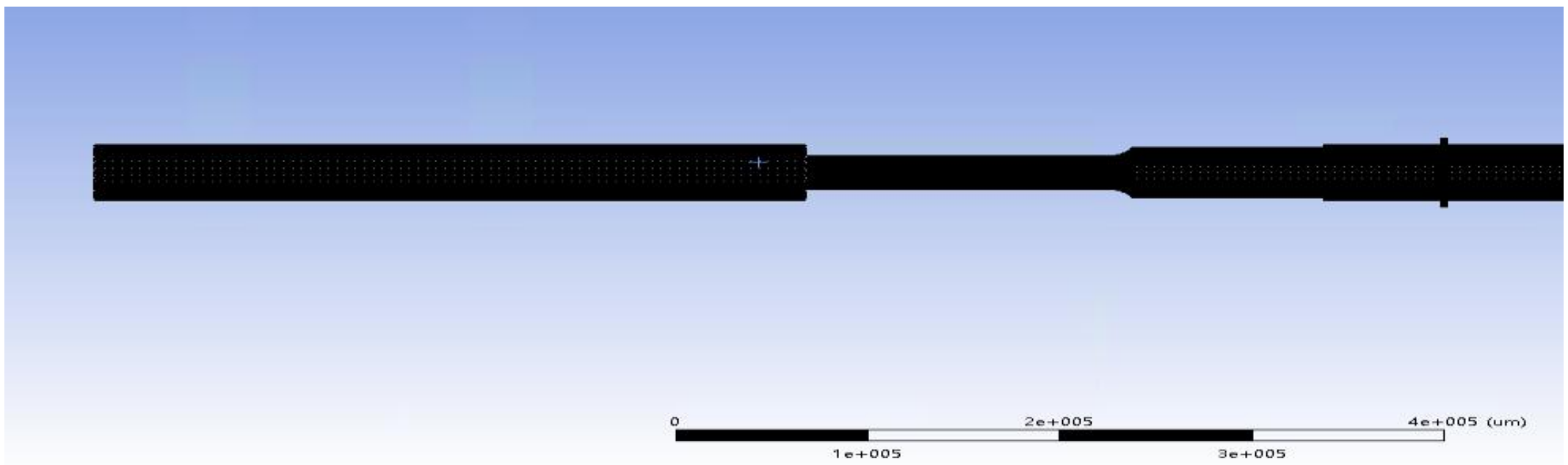

(b) Grid division diagram of radiation rod after surface treatment

**Fig. 3. Mesh division diagram of ultrasonic vibration system**

(3) Model loading and solving

Modal analysis: the positive and negative poles of the piezoelectric slice should be coupled with voltage degrees of freedom, and the frequency range of modal extraction was set as 18~ 22kHz. The fixed support plane was selected as the plane of the piezoelectric ceramic rear end cover.

Harmonic response analysis: the harmonic response analysis frequency range is set as 20~20.6KHz based on the natural frequency obtained from the modal analysis, and the bolt pretightening force shall be added to the front end cover of the piezoelectric ceramics.

The vibration direction and polarization direction of the piezoelectric ceramic plate are set as the Y axis, and then the excitation voltage with the amplitude of U=390V is added on the surface of the piezoelectric ceramic plate, and the voltage at the top free surface is U=0V. The pretightening bolt force generated by the pretightening bolt at the upper end of the piezoelectric ceramic sheet affects the dynamic characteristics of the ultrasonic vibration system, and the pretightening bolt force is set as 83.4N[13-16].

**2.2Simulation of sound pressure field in aluminum melt**

(1) Establishment of finite element model of fluid region

The sound field propagates in the medium. According to bounded or unbounded, the distribution of the sound field is divided into near field and far field. In the simulation of ultrasonic sound field distribution, the unbounded area is generally selected for the acoustic analysis of simulation. For laboratory analysis of ultrasonic casting system, sound wave propagates in aluminum melt, and it is generally difficult to determine the distribution of sound field due to the absorption and reflection characteristics of sound wave on crucible wall. Therefore, this paper adopts unbound region to simulate the sound field [17-20].

Figure 4-(a) and 4-(b) are respectively the schematic diagram of the ultrasonic casting system used in the experiment of this research group and the meshing diagram in simulation. The ultrasonic vibration system is selected the same as above, the size of ultrasonic radiation rod ∅50×350mm, and the radiation rod with ion nitriding and vacuum carburizing thickness of 55μm and 53μm respectively after surface treatment is selected. In order to facilitate the observation and selection of 1/2 section of the model, the diameter of the molten pool D=630mm. In the three cases, the vibration position is at the center of the molten pool, perpendicular to the molten pool, with a depth of 280mm. The resonant frequency of the ultrasonic radiation rod is 20KHz. Due to the working principle and action principle of the ultrasonic vibration system, three fields of piezoelectric, acoustic and structure are required to be coupled during the simulation[15]. The material properties of the ultrasonic vibration system are still consistent with the previous dynamic simulation analysis, and the parameters of the required components are shown in Table 2.

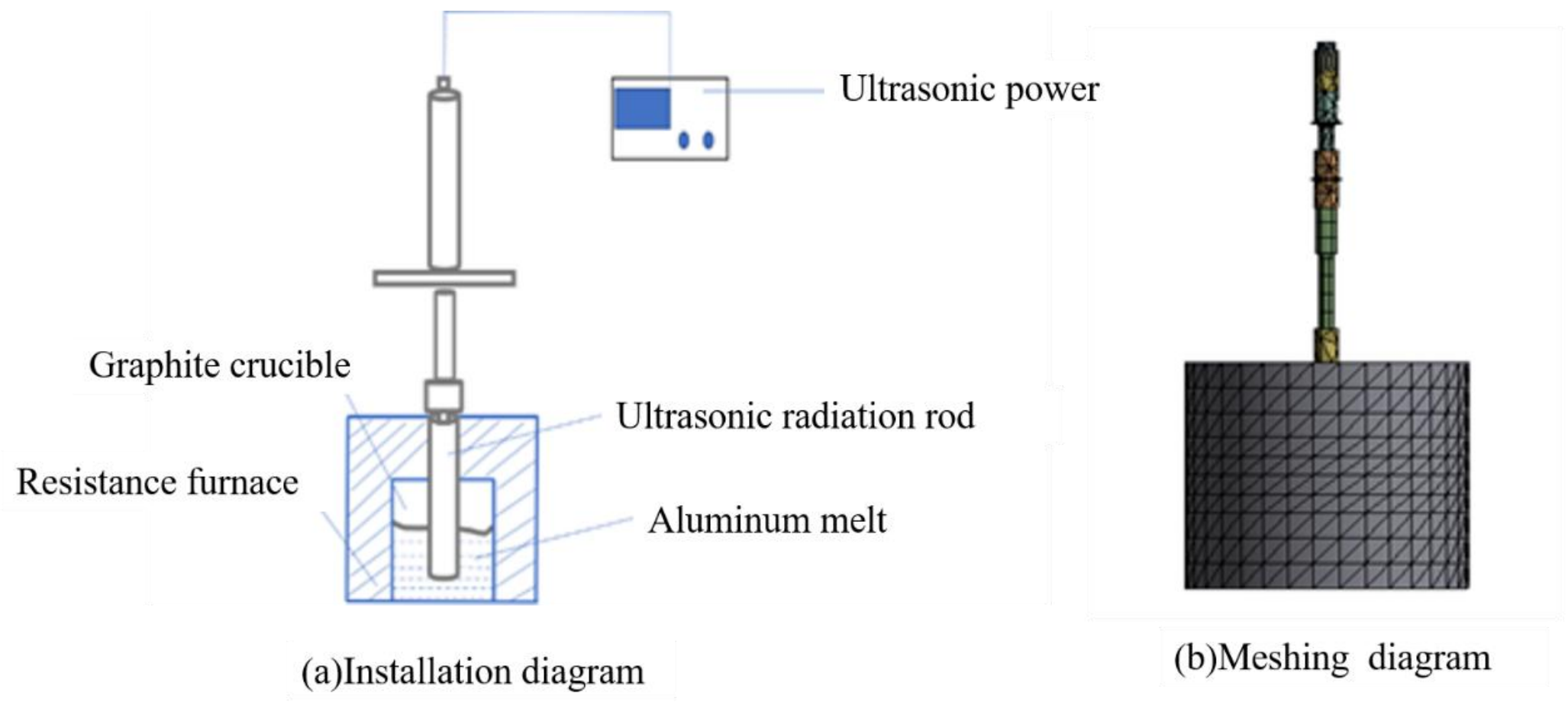

**Fig. 4.** Schematic diagram and mesh division diagram of ultrasonic casting system

(2) Boundary conditions and model loading

Setting of sound field boundary: the side and bottom of the molten pool are set as absorption surfaces, the surface where the aluminum melt contacts with the radiation rod is a free interface, and the structure-fluid coupling boundary conditions are applied[14]. The excitation voltage U=390V on the surface of the piezoelectric ceramic plates is used as the positive electrode, and the voltage U=0V on the top free surface is used as the negative electrode. The analysis type was selected as acoustic pressure harmonic response analysis, and the frequency range was set as 20kHz ~ 20.6kHz. The material properties of aluminum melt are shown in Table 2.

### 2.3 The simulation results

(1) Analysis of dynamic characteristics of ultrasonic radiation rod

The dynamic simulation results, modal analysis and harmonic response analysis results of the three kinds of ultrasonic radiation rods are shown in Fig. 5 and Fig. 6.

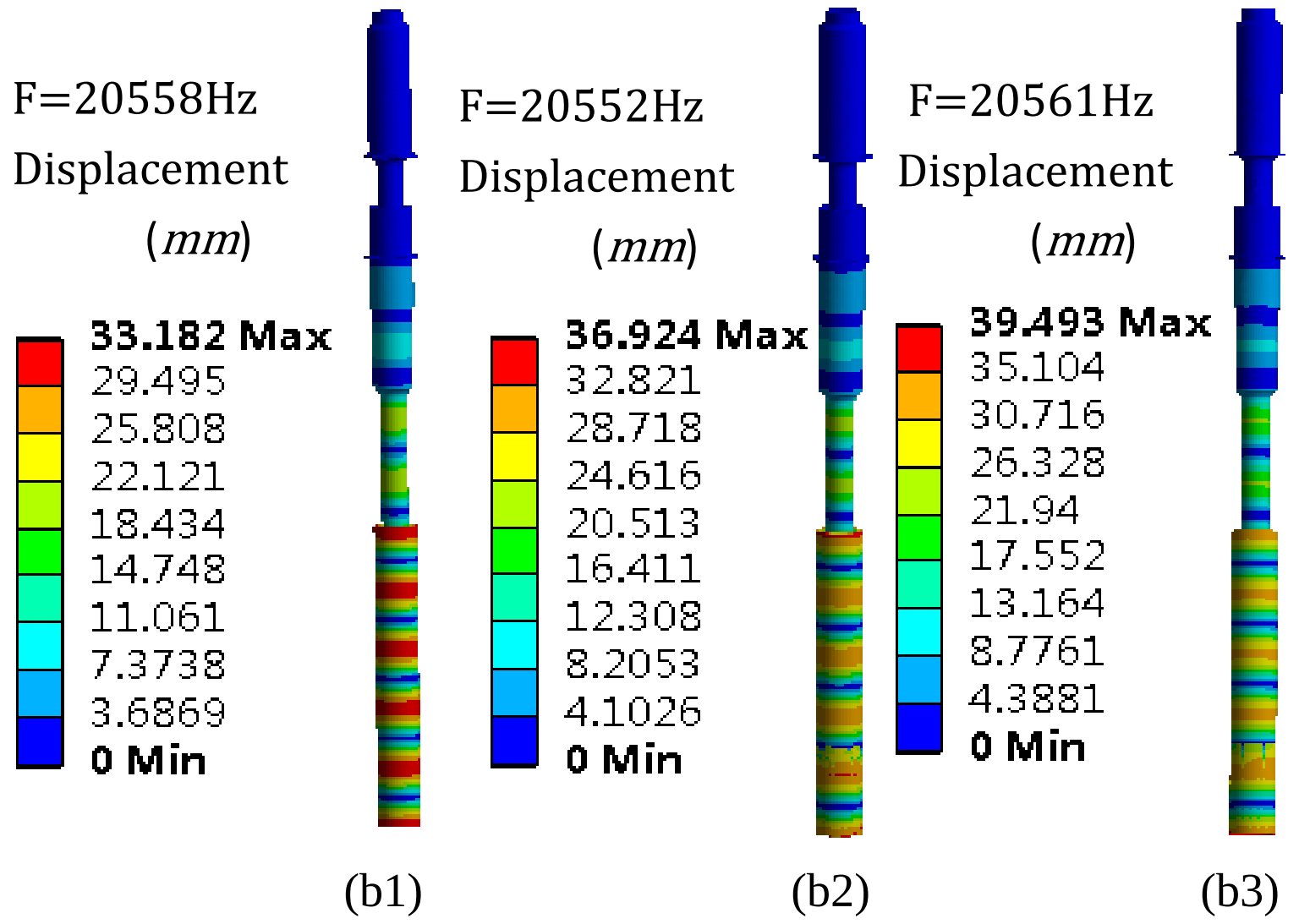

(b1)-(b3) Untreated, vacuum carburizing, ion nitriding

**Fig. 5.**Modal analysis diagram of three kinds of ultrasonic radiation rods

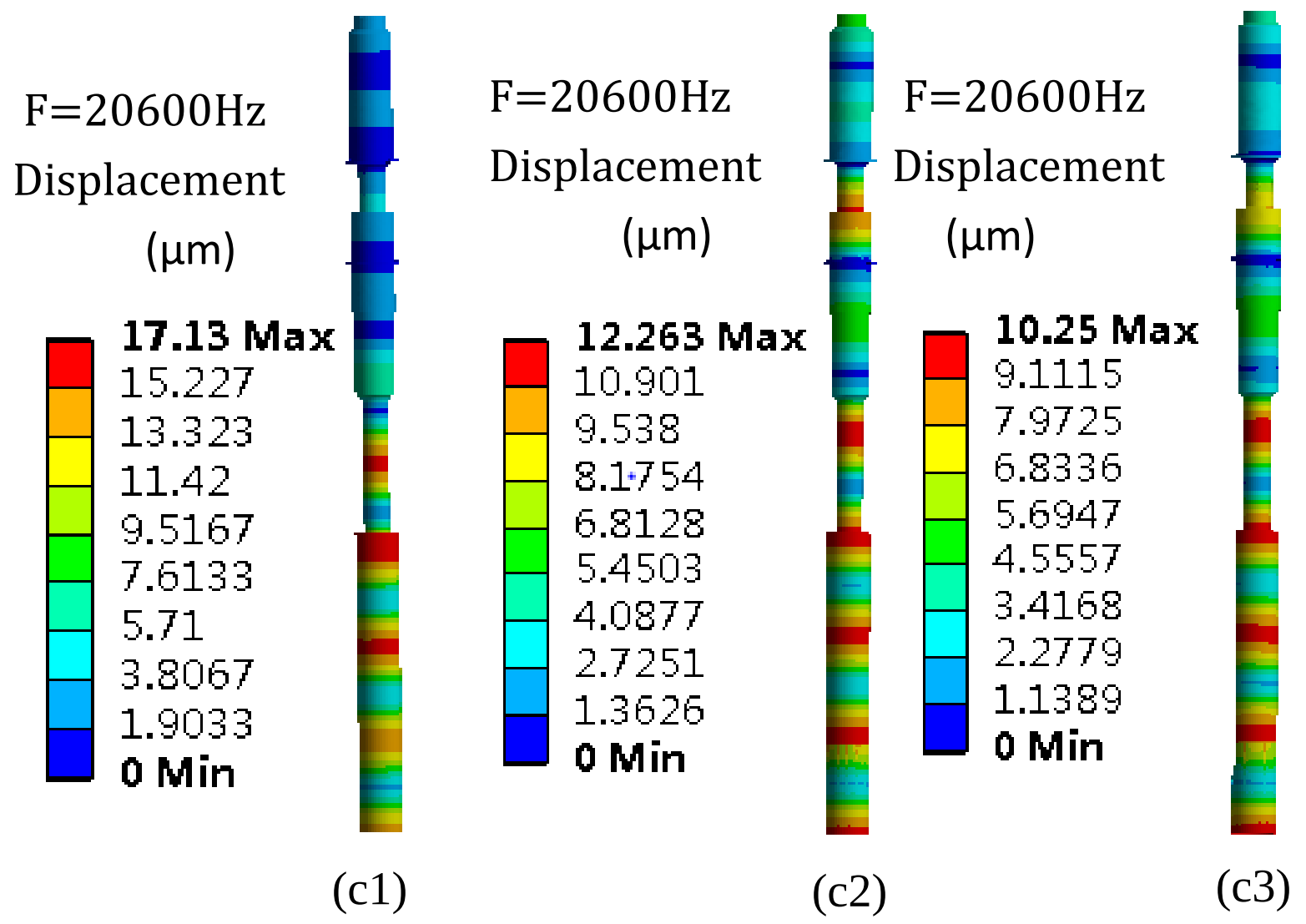

(c1)-(c3) Untreated, vacuum carburizing, ion nitriding

**Fig. 6.**Harmonic response analysis diagram of three radiation rods

① Modal analysis

The natural frequencies of three kinds of ultrasonic radiation rods from 19 kHz to 22 kHz were extracted. According to the modal simulation result 5- (b1-b3), the natural frequencies of three kinds of ultrasonic vibration systems were respectively 20558Hz, 20561Hz and 20552Hz, and the ultrasonic radiation rods were mainly longitudinal vibration. It can be seen that the radiation rod after surface treatment has little effect on the natural frequency of the ultrasonic vibration rod system.

② Harmonic response analysis

The modal superposition of the three radiation rods is performed to obtain the vibration displacement at the natural frequency. The simulation results of harmonic response analysis are shown in Fig. 6- (c1-c3). It can be seen from Fig. 6 that the end face amplitudes of the three radiation rods are the largest, and the end face amplitudes are 17.13μm, 13.23μm and 10.25μm respectively. The longitudinal vibration displacement of the ultrasonic radiation rod shows a trend of cosine vibration with the distance from the end of the radiation rod. After the surface treatment, the amplitude of the end face of the ultrasonic radiation rod decreased by 3.9 μm and 6.88 μm, respectively, and the amplitude of the end face of the radiation rod treated by nitriding was the smallest.

(2) Distribution of sound pressure field in aluminum melt

The acoustic pressure harmonic response analysis results of ultrasonic radiation rod aluminum melt ultrasonic casting system with different surface treatments (untreated, vacuum carburizing and ion nitriding) are shown in Fig.7. Through comparison, it can be found that the acoustic pressure distribution field of ultrasonic radiation rod treated by vacuum carburizing in aluminum melt is similar to that of untreated and ion nitriding. The sound pressure field is distributed symmetrically with the radiation rod as the center. Due to the large ultrasonic vibration energy of the end face of the ultrasonic radiation rod, the sound pressure amplitude near it is the largest. With the increase of the distance from the end face of the ultrasonic radiation rod, the sound pressure amplitude decreases rapidly, and there are alternating positive and negative pressures. A small area of low amplitude sound pressure appears on the side of the radiation rod. But the amplitude of sound pressure in the three kinds of radiation rod aluminum melt is obviously different. The amplitude of sound pressure at the end of radiation rod is 4.3MPa, 3.6MPa and 3.3MPa, respectively. As can be seen from Fig. 8, the change trend of sound pressure in aluminum melt and the comparison of sound pressure amplitude in three kinds of radiation rod aluminum melt show that the sound pressure amplitude of ultrasonic radiation rod in aluminum melt after surface treatment is significantly reduced, and the sound pressure amplitude generated by nitroded ultrasonic radiation rod in aluminum melt is the least.

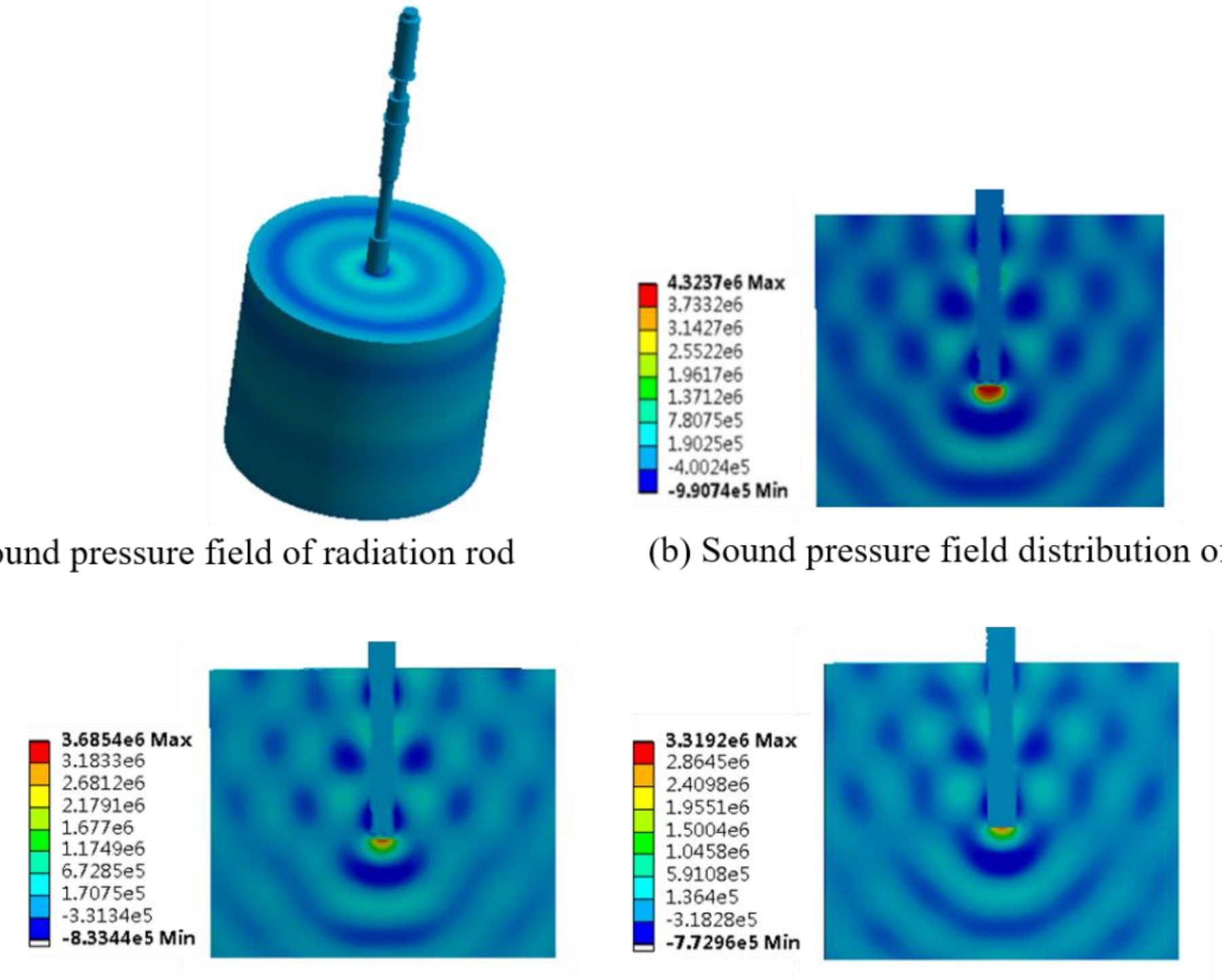

(a) Sound pressure field of radiation rod (b) Sound pressure field distribution of untreated radiation rod

(c) Sound pressure field distribution of vacuum carburizing radiation rod (d) Sound pressure field distribution of ion nitriding radiation rod

**Fig. 7.** 1/2 cross section of sound pressure simulation results of three radiation rods

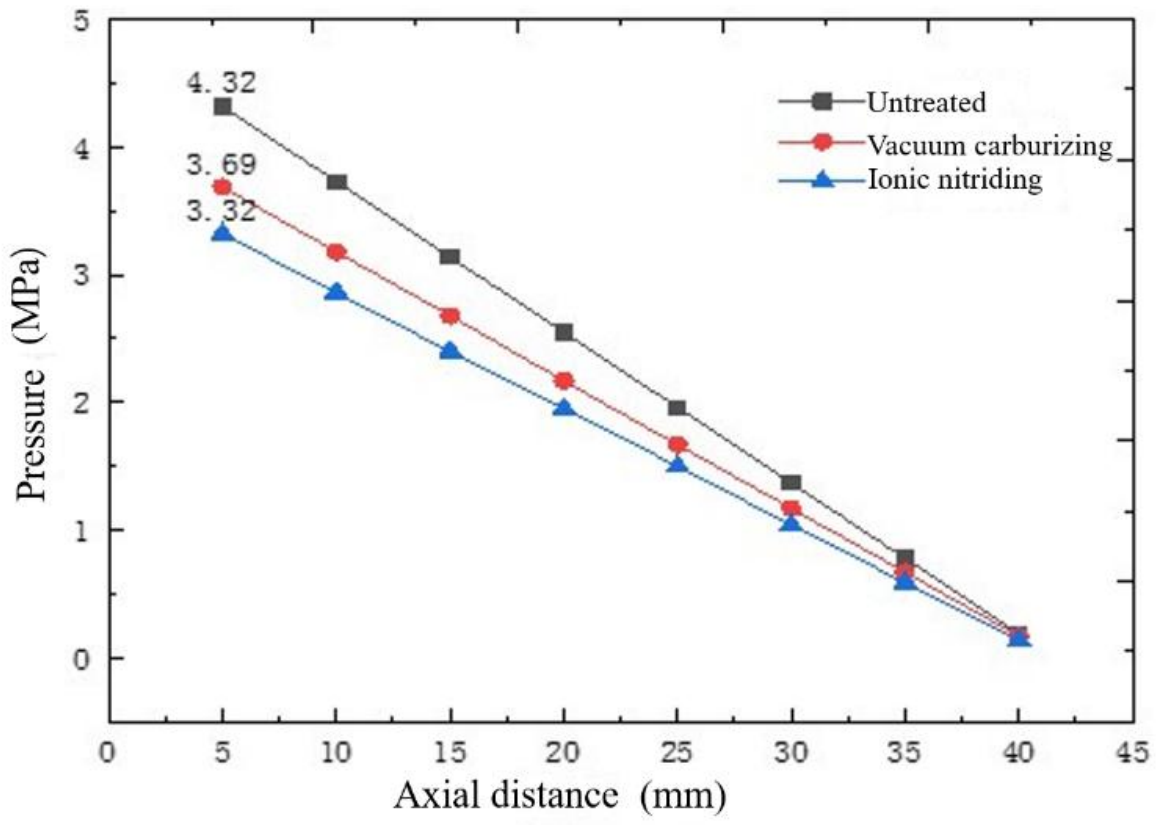

**Fig. 8.** Distribution of sound pressure field on cylinder surface of three ultrasonic casting systems

## 3.Estimation and experiment of cavitation field generated by three kinds of ultrasonic radiation rods under ultrasonic vibration

The effect of power ultrasound depends on the cavitation domain generated by the cavitation effect, and the distribution of the cavitation domain changes with the change of the sound pressure field[21-22]. Due to the high temperature and closed invisibility in the process of ultrasound-assisted casting, it is impossible to observe directly. In order to more intuitively observe the distribution of ultrasonic sound pressure field in the medium and compare it with the simulation results of the model. Therefore, in this chapter, ultrasonic radiation rods with different surface treatments were used to conduct aluminum foil cavitation erosion water experiment, and the distribution of cavitation domain in water was observed based on the results of aluminum foil cavitation erosion, so as to speculate the influence of the change of sound pressure amplitude of the three radiation rods in aluminum melt on the formation of ultrasonic radiation rod vibration on cavitation region.

### 3.1 Experimental equipment and scheme

The experiment uses water as liquid medium and pure aluminum foil as corrosive material. The experimental equipment mainly includes: ultrasonic vibration system (ultrasonic transducer, amplitude transformer, radiation rod) radiation rod including untreated and carburizing, nitriding treatment of radiation rod. The radiation rod is made of titanium alloy (Φ50× 380mm), and the ultrasonic power supply (18-22kHz) can track the resonant frequency in real time and adjust the voltage. Auxiliary equipment includes: beaker, adjustable support frame, hardboard, scotch tape, HD camera.

Put an appropriate amount of water into the beaker and place it on the workbench.

Cut out the hard plate so that it can be put into the beaker. Cut out the concave hard plate according to the insertion depth of the radiation rod so that the radiation rod can be just inserted. The aluminum foil is also cut in the same way and fixed on both sides of the board with transparent tape. The ultrasonic radiation rod was inserted into the water, and the support frame was adjusted so that the depth of the submerged liquid surface was 280mm. The ultrasonic power supply was connected to the ultrasonic vibration system, and the frequency of the power supply was adjusted to 20kHz, the voltage was 380V, the power was 600W, and the ultrasonic vibration time was 20s. Turn off the ultrasonic power, take the aluminum foil out of the water and dry it to observe the distribution of the corrosion holes over the filter. According to the above process, the untreated, carburized and nitriding radiation rods were used for experiments respectively. The schematic diagram and physical diagram of the experimental device are shown in Fig. 9.

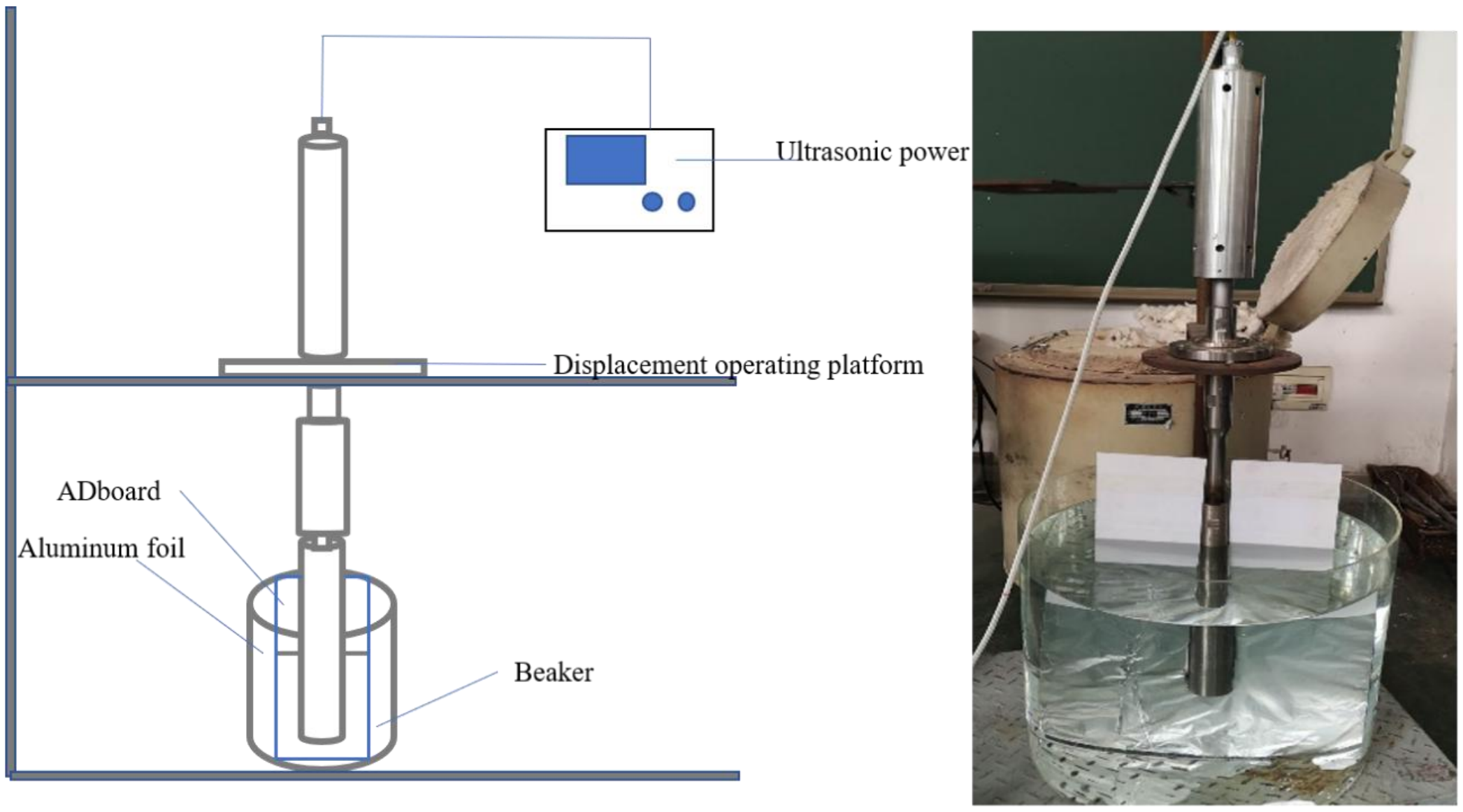

（a）Schematic diagram of experimental device

(b) Physical picture of experiment

**Fig.9.**Schematic diagram and physical diagram of the experimental device for aluminum foil cavitation erosion in water

## 3.2 Aluminum foil cavitation corrosion test results

Fig.10 is the experimental diagram of aluminum foil cavitation erosion in water of three kinds of ultrasonic radiation rods. From the macro morphology of aluminum foil cavitation corrosion, it can be observed that the surface of aluminum foil has different degrees of corrosion, and cavitation exists in a specific area.

Fig.10 (a) shows the macroscopic distribution of aluminum foil corrosion when untreated radiation rod is used. It can be seen that cavitation corrosion is mainly concentrated in the range of about 40mm axial and 20mm radial under the end face of radiation rod. Fig.(a1) is a local enlarged view of the cavitation area. It can be seen that the depth and size of the pits are relatively uniform with the increase of the axial distance.

Fig.10 (b) shows the macroscopic distribution of aluminum foil corrosion in water when the radiation rod treated by carburizing is used. It can be seen that the cavitation erosion area is mainly distributed in the range of about 35mm axial and 25mm radial under the end face of the radiation rod, and the cavitation holes in this area are the most densely distributed. As can be seen from the local enlarged figure (b1), the pitting holes are dispersed and the pitting depth and size are not uniform.

Fig.10 (c) shows the macroscopic distribution of aluminum foil corrosion when nitriding radiation rod is used. It can be seen that the cavitation erosion area is mainly concentrated in the range of about 30mm axial and 15mm radial of the end face of the radiation rod. Through comparison, it can be found that the surface treated ultrasonic radiation rod aluminum foil cavitation erosion area is obviously reduced, and the number of pitting holes is small, the depth of pitting holes is shallow. Among the two surface treatments, the cavitation erosion area and the degree of cavitation erosion are the least in the aluminum foil under the radiator rod after ion nitriding.

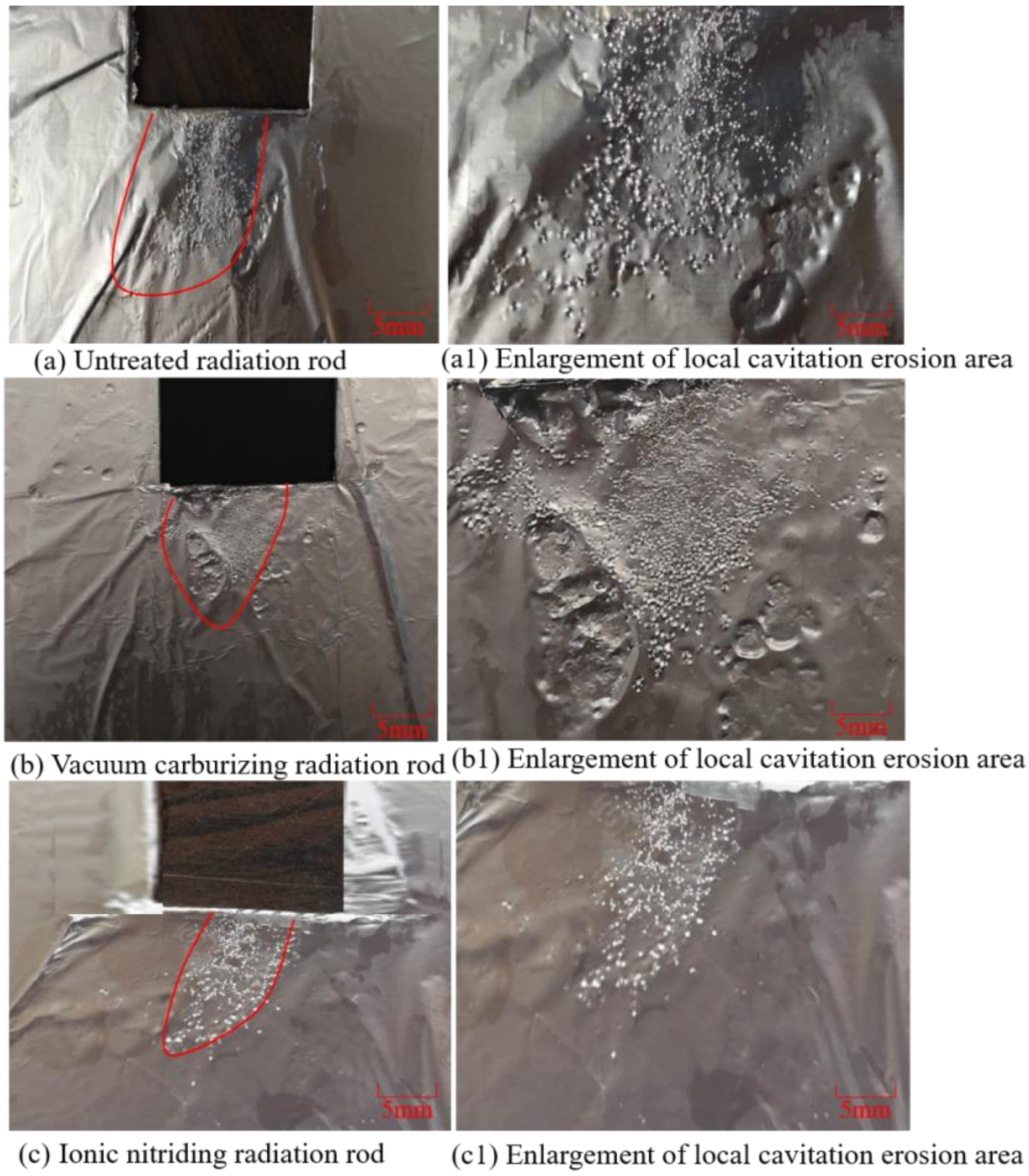

**Fig. 10**.Macro morphology of aluminum foil cavitation etching in water experiment with three kinds of ultrasonic radiation rods

## 4.Comparison and analysis of results

(1) Analysis of simulation results

According to the results of modal analysis and harmonious response analysis, the amplitude of the end face of the ultrasonic radiation rod decreases by 33% and 46% respectively after the surface treatment, which will lead to the decrease of the ultrasonic power transferred to the aluminum melt. According to the simulation results of sound pressure field in aluminum melt, the sound pressure amplitude of ultrasonic radiation rod in aluminum melt after surface treatment is obviously lower, which indicates that the ability of ultrasonic radiation rod to transfer ultrasonic power is weakened after surface treatment, and the sound pressure value in aluminum melt is reduced. This is consistent with the changing trend of the vibration amplitude of the ultrasonic radiation rod. The surface treatment increases the surface elastic modulus of the radiation rod, increases the vibration damping of the radiation rod, and suppress the vibration of the ultrasonic radiation rod, which leads to the decrease of the ultrasonic energy transmitted to the aluminum melt, and then affects the sound pressure value in the aluminum melt. Moreover, the greater the elastic modulus is, the more obvious the vibration suppression effect is. Therefore, the amplitude of sound pressure generated at the end of the nitroded ultrasonic radiation rod is reduced by about 23% compared with that of the untreated ultrasonic radiation rod, and the vibration suppression effect of the radiation rod is the most obvious, making the amplitude of sound pressure generated in the aluminum melt the least.

(2) Analysis of experimental results

The three kinds of ultrasonic radiation rod aluminum foil cavitation erosion experiments respectively, can be found using the lever surface treatment of radiation experiments the cavitation erosion area of the aluminum foil is lower than with untreated radiation bar of aluminum foil cavitation erosion area, axial cavitation erosion area reduced about 12.5% respectively, 25%, using ion nitriding treatment of radiation rod, minimum cavitation erosion experiment of aluminum foil area. When the sound pressure value generated by the ultrasonic vibration system in the aluminum melt is greater than or equal to 1.1MPa, the collapse of cavitation bubbles in the negative pressure state will lead to the occurrence of cavitation phenomenon in the aluminum melt [23-24]. The decrease of sound pressure value will lead to the decrease of the number of cavitation bubbles and the impact force of micro-jet generated by the collapse of cavitation bubbles, and the damage degree to the aluminum foil surface will be weakened. The results show that the acoustic pressure generated by the surface treated ultrasonic radiation rod in the aqueous solution is small, which affects the cavitation domain and the depth of the pitting hole.

(3) Comparison of Results

Comparison of sound pressure simulation results and aluminum foil cavitation corrosion experiment is shown in Fig. 11. Approved by contrast can be found that the surface treatment of radiation aluminum foil cavitation erosion area reduced obviously, the axial cavitation erosion area reduced about 12.5%, respectively, 25%. The size of cavitation erosion area reflects the size of sound pressure amplitude, that is, the results of aluminum foil cavitation erosion experiment in water are similar to those of sound pressure amplitude simulation, and cavitation holes are gradually sparse with the increase of axial distance, indicating that the sound pressure amplitude is gradually decreasing, which is the same as the distribution rule of sound pressure amplitude obtained by sound pressure simulation. The sound pressure of surface treated ultrasonic radiation rod in aqueous solution is smaller than that of untreated ultrasonic radiation rod, and the sound pressure of ion nitriding radiation rod is the least. The rationality of simulation results of sound pressure field is verified by experiments.

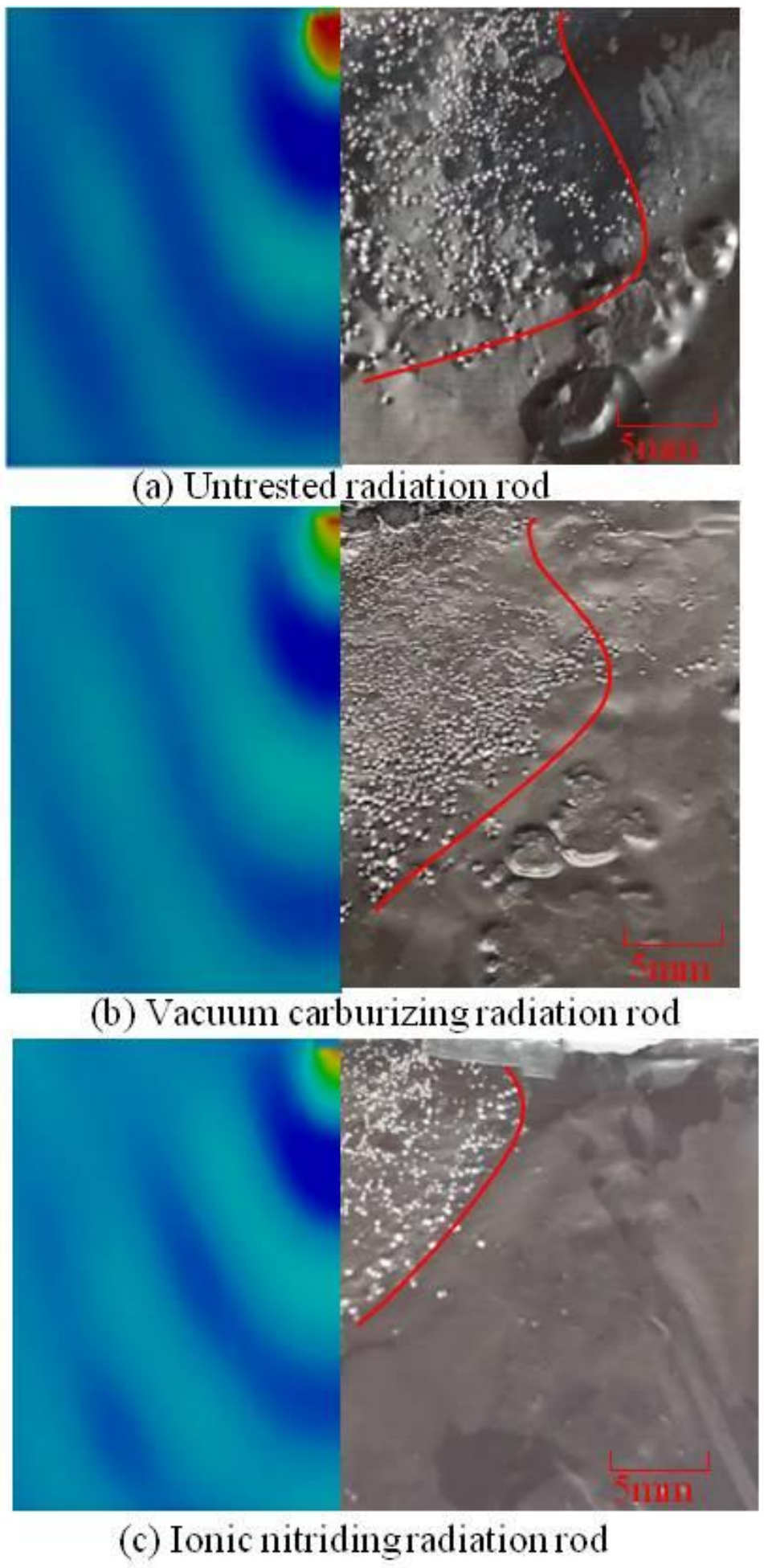

**Fig. 11.**Comparison between sound pressure simulation and aluminum foil cavitation etching morphology

## 5.Conclusion

(1) Two kinds of surface treatment can reduce the amplitude of u ultrasonic radiation rod by 4.9μm and 6.9μm, respectively. The acoustic pressure amplitude of radiation rod in aluminum melt decreases by 0.7 MPa and 1 MPa, respectively. The acoustic pressure amplitude of radiation rod treated by ion nitriding in aluminum melt is the lowest.

(2) After surface treatment, the elastic modulus and hardness of the radiation rod surface increase, the vibration response of the radiation rod is weakened, and the amplitude of the radiation rod is reduced. The larger the surface elastic modulus of the radiating rod is, the more obvious the inhibition effect on the vibration of the ultrasonic radiating rod is.